\documentstyle[epsfig,12pt]{article}  
\newcommand{\beq}{\begin{equation}}
\newcommand{\eeq}{\end{equation}}
\newcommand{\bea}{\vspace{0.25cm}\begin{eqnarray}}
\newcommand{\eea}{\end{eqnarray}}

\newcommand{\r}{\mbox{{\boldmath
$\rho$}}}

\newcommand{\qb}{\mbox{{\bf
q}}}

\def\lsim{\mathrel{\rlap{\lower4pt\hbox{\hskip1pt$\sim$}}
    \raise1pt\hbox{$<$}}}         %less than or approx. symbol
\def\gsim{\mathrel{\rlap{\lower4pt\hbox{\hskip1pt$\sim$}}
    \raise1pt\hbox{$>$}}}         %greater than or approx. symbol
\begin{document}
\thispagestyle{empty}
\vspace*{-2cm}
 
\bigskip

\begin{center}

  {\Large\bf
Induced photon emission from quark jets in 
ultrarelativistic heavy-ion collisions
 \\
\vspace{1.5cm}
  }
\medskip
{\large
  B.G.~Zakharov
  \bigskip
  \\
  }
{\it  
 L.D.~Landau Institute for Theoretical Physics,
        GSP-1, 117940,\\ Kosygina Str. 2, 117334 Moscow, Russia
\vspace{2.7cm}\\}

  {\bf
  Abstract}
\end{center}
{
\baselineskip=9pt
We study the induced photon bremsstrahlung from a fast quark 
produced in $AA$-collisions due to multiple scattering in quark-gluon plasma.
For RHIC and LHC conditions the induced photon spectrum is sharply 
peaked at photon energy close to the initial quark energy. In this region
the contribution of the induced radiation to the photon fragmentation 
function exceeds the ordinary vacuum radiation.
Contrary to previous analyses \cite{DH,JOS2,JOS1,JJS} 
our results show that at RHIC and LHC energies the final-state 
interaction effects in quark-gluon plasma do not suppress the direct 
photon production, and even may enhance it at $p_{T}\sim 5-15$ GeV.\\
}
%\pagebreak
%-------------------------------------------------------------
\bigskip\\
\noindent{\bf 1}. In resent years much attention has been attracted to the
direct photon production in $AA$-collisions at RHIC and LHC energies
(see, for example, \cite{A,PT,CERN} and references therein).
It is expected that for sufficiently small transverse momenta 
($p_{T}\lsim 3-4$ GeV) 
the dominating source of the direct photons at RHIC and LHC 
is radiation from
quark-gluon plasma (QGP), and at higher $p_{T}$ it is hard partonic mechanisms
(Compton-process, 
quark-antiquark annihilation,
and bremsstrahlung from fast quarks (antiquarks)
produced in hard reactions) \cite{A,PT}. 
Nuclear effects should modify the pQCD partonic contribution to the direct 
photons in $AA$-collisions as compared to that in $pp$ interaction.
For the Compton and annihilation processes which occur at small distances 
this modification, related to 
the initial state interaction (ISI) effects 
(nuclear shadowing and Cronin effect),
is relatively small. However, one can expect strong final-state 
interaction (FSI) effects in the QGP for the bremsstrahlung 
which occurs at large distances.
Investigation of the influence of the FSI on photon bremsstrahlung is 
of great interest from the point of view using the direct photons as a probe
for the QGP at RHIC and LHC. It is especially important for LHC energies
where the bremsstrahlung component is the dominating partonic mechanism
\cite{A,DH,JOS2}. 

It was suggested \cite{DH,JOS2,JOS1,JJS} that at RHIC and LHC 
the quark energy loss in the QGP phase due to the induced gluon radiation
related to the multiple scattering should suppress strongly 
the bremsstrahlung contribution. At LHC energies it is equivalent to
strong suppression of the total contribution of the pQCD mechanisms.
In \cite{DH,JOS2,JOS1,JJS} it was assumed that the only effect of QGP on photon
bremsstrahlung comes from the shift of the quark energy to a smaller value
due to gluon emission before photon radiation.
However, the analyses \cite{DH,JOS2,JOS1,JJS} missed two essential points. 
First of all, multiple scattering 
which fast quarks undergo in the 
QGP must enhance photon radiation due to the induced photon 
emission.
Also, the assumption that all gluons are radiated before photon emission
is not justified since the formation
lengths for gluon and photon radiation are of the same order. 
In this case the gluons radiated after the photon do not suppress 
photon emission. For this reason one can expect that analyses 
\cite{DH,JOS2,JOS1,JJS} overestimate suppression of photon bremsstrahlung. 
In the present paper we address the photon bremsstrahlung at 
RHIC and LHC energies accounting for the induced photon radiation
and the effect of finite gluon formation length.
%For LHC energies where the the bremsstrahlung component is the dominating
%partonic mechanism our results give an estimate for the direct photon
%production.

\vspace{.2cm}
\noindent{\bf 2}. The contribution of the bremsstrahlung mechanism to 
the cross section of photon production can be written as \cite{Owens,ABFS}
\beq
\frac{d\sigma^{AA}_{\gamma}(p_{T})}{dyd p_{T}^{2}}=
\int\limits_{0}^{1}\frac{dx}{x^{2}} D_{q\rightarrow \gamma}(x,p_{T}/x)
\frac{d\sigma^{AA}_{q}(p_{T}/x)}{dydp_{T}^{2}}\,,
\label{eq:1}
\eeq
where
${d\sigma^{AA}_{q}(p_{T})}/{dydp_{T}^{2}}$
is the cross section of the processes $A+A \rightarrow q+X$
(hereafter we suppress the argument $y$ and mean the central rapidity
region $y\approx 0$),
$D_{q\rightarrow \gamma}(x,E)$ is the fragmentation function for 
the $q\rightarrow \gamma q$
transition which accounts for all the FSI effects
($x=E_{\gamma}/E$, $E_{\gamma}$ and $E$ is the photon and 
initial quark energies, respectively).
Summation over quark (antiquark) states is implicit on the right-hand side
of (\ref{eq:1}). 
We write $D_{q\rightarrow \gamma}(x,E)$ in the form 
\beq
D_{q\rightarrow \gamma}(x,E)=
\frac{dP_{vac}(x,E)}{dx}+\frac{dP_{ind}(x,E)}{dx}\,.
\label{eq:1p}
\eeq
The first term on the right-hand side of (\ref{eq:1p}) is the probability 
distribution of the ordinary vacuum $q\rightarrow \gamma q$ splitting and 
the second one
corresponds to the induced transition due to quark multiple scattering.

An accurate evaluation of the induced photon emission from a 
fast quark requires 
treatment of photon and multiple gluon radiation on an even footing. 
We begin with discussion of the induced $q\rightarrow \gamma q$ 
transition
ignoring gluon radiation. A qualitative method for accounting for the gluon
effects will be discussed later.
To evaluate the induced spectrum
we use the light-cone path integral approach \cite{Z1} (see also
\cite{Z2,Z_YAF,Z3,BSZ})
which allows one to account for the Landau-Pomeranchuk-Migdal (LPM) effect
\cite{LP,Migdal} and finite-size effects which play an important 
role in the problem of interest.
The induced photon spectrum for a fast quark produced at $z=0$ can 
be written in a form similar to that for gluon emission given in \cite{Z4}
(we take $z$ axis along the quark momentum) 
\beq
\frac{d P_{ind}}{d
x}=
\int\limits_{0}^{\infty}\! d z\,
n(z)
\frac{d
\sigma_{eff}^{BH}(x,z)}{dx}\,,
\label{eq:3}
\eeq
where $n(z)$ is the number density of 
the medium, and
\beq
\frac{d
\sigma_{eff}^{BH}(x,z)}{dx}=\mbox{Re}
\int d\r\,
\Psi^{*}(\r,x){\sigma}(\rho x)\Psi_{m}(\r,x,z)\,
\label{eq:4}
\eeq
is the in-medium ($z$-dependent) Bethe-Heitler cross section.
Here ${\sigma}(\rho)$ is the dipole cross section of a
quark-antiquark pair of size $\rho$ with a particle in the medium. 
$\Psi(\r,x)$ is the 
light-cone wave function for the $q\rightarrow \gamma q$ transition in vacuum, 
and $\Psi_{m}(\r,x,z)$ is the in-medium light-cone wave function at 
the longitudinal coordinate $z$ (we omit spin indices). 
The no spin flip wave functions, dominating the spectrum, read 
\bea
\Psi(\r,x)&=& 
P(x)\left (\frac{\partial}{\partial \rho_{x}^{'}}
-i s_{g}\frac{\partial}{\partial \rho_{y}^{'}}\right )
\int\limits_{0}^{\infty}d\xi
\exp\left(-\frac{i\xi}{L_{f}^{\gamma}}\right)
{\cal{K}}_{0}(\r,\xi|\r^{'},0)\Biggl |_{\r^{'}=0}\,,
\label{eq:5}\\
\Psi_{m}(\r,x,z)&=&
P(x)\left (\frac{\partial}{\partial \rho_{x}^{'}}
-i s_{g}\frac{\partial}{\partial \rho_{y}^{'}}\right )
\int\limits_{0}^{z}d\xi
\exp\left(-\frac{i\xi}{L_{f}^{\gamma}}\right)
{\cal{K}}(\r,z|\r^{'},z-\xi)\Biggl |_{\r^{'}=0}\,,
\label{eq:6}
\eea
where $P(x)=
i e_{q}\sqrt{{\alpha_{em}}/{2x}}
[s_{\gamma}(2-x)+2s_{q}x]/2M(x)$  ($s_{q,\gamma}$ denote quark and photon 
helicities),
$
L_{f}^{\gamma}={2E(1-x)}/{m_{q}^{2}x}
$ is the photon formation length,
${\cal{K}}$ is the Green's function
for the two-dimensional Hamiltonian
\beq
\hat{H}(z)=-\frac{1}{2M(x)}
\left(\frac{\partial}{\partial \r}\right)^{2}
          -i\frac{n(z){\sigma}(\rho x)}{2}\,
\label{eq:7}
\eeq
with $M(x)=Ex(1-x)$, and 
\beq
{\cal{K}}_{0}(\r_{2},z_{2}|\r_{1},z_{1})=\frac{M(x)}{2\pi i(z_{2}-z_{1})}
\exp\left[\frac{iM(x)(\r_{2}-\r_{1})^{2}}{2(z_{2}-z_{1})}\right]
\label{eq:8}
\eeq
is the Green's function for the Hamiltonian (\ref{eq:7}) with $n(z)=0$.

The dipole cross section reads 
\beq
\sigma(\rho)=C(\rho)\rho^{2}\,,
\label{eq:9}
\eeq
where
\beq
C(\rho)=\frac{C_{T}C_{F}}{\rho^{2}}\int d\qb\alpha_{s}^{2}(q^{2})
\frac{[1-\exp(i\qb\r)]}{(q^{2}+\mu^{2}_{D})^{2}}\,\,.
\label{eq:10}
\eeq
Here $C_{T,F}$ is the color Casimir for the medium constituents 
(quarks and gluons) and quark, $\mu_{D}$ is the Debye screening mass.

The effective Bethe-Heitler cross section (\ref{eq:4}) differs from that for 
a quark incident from infinity on an isolated scattering center
due to the LPM effect and finite-size effects originating from
the in-medium light-cone wave function entering (\ref{eq:4}). In the 
high-energy limit,
when $L\ll L_{f}^{\gamma}$, here $L$ is the thickness of the medium, 
the typical values of $\rho$ become small 
($\sim \sqrt{2L/Ex(1-x)}$), and the LPM effect can be neglected. 
In this case
the spectrum is dominated by the $N\!=\!1$ scattering.  
In this regime the most important effect is 
modification of the in-medium light-cone wave function due to the 
finite-size effects. Let us first discuss the $N\!=\!1$ term
to illustrate qualitatively the role of the finite-size effects.
Neglecting the $Q^{2}$-dependence of $\alpha_{s}$ from 
(\ref{eq:4})-(\ref{eq:10}) one can obtain
\beq
\frac{d
\sigma_{eff}^{BH}(x,z)}{dx}=\frac{\pi\alpha_{em}\alpha_{s}C_{T}C_{F}}
{4E}\cdot\frac{(1-x+x^{2}/2)z}{(1-x)}
\label{eq:11}
\eeq
(hereafter for simplicity it is implied that $e_{q}=1$).
The derivation of (\ref{eq:11}) is similar to that for gluon 
emission given in \cite{Z_gluon}.
After substitution of (\ref{eq:11}) in (\ref{eq:3}) one obtains 
(for $n(z)=$const) the spectrum $\propto L^{2}$
\beq
\frac{dP_{ind}(x,E)}{dx}=\frac{\pi\alpha_{em}\alpha_{s}C_{T}C_{F}nL^{2}
(1-x+x^{2}/2)}{8E(1-x)}\,.
\label{eq:12}
\eeq
One sees from (\ref{eq:12}) that in the high-energy limit contrary to 
the ordinary
Bethe-Heitler spectrum $\propto 1/x$  the bremsstrahlung is $\propto 1/(1-x)$.
Of course, formulas (\ref{eq:11}), (\ref{eq:12}) become invalid at 
$(1-x)\ll Lm_{q}^{2}/E$ when 
$L_{f}^{\gamma}\ll L$ and the spectrum reduces to the ordinary 
Bethe-Heitler one.
It is worth noting that in the diagrammatic language the above formula for 
$N\!=\!1$ term corresponds to the set of 
diagrams shown in Fig.~1. 
We would like to emphasize that the spectrum (\ref{eq:12}), 
similarly to the gluon spectrum of Ref. \cite{Z_gluon}, cannot be obtained
if one ignores the logarithmic $\rho$-dependence of the function 
$C(\rho)$ (\ref{eq:10}) at small $\rho$. One can see that, similarly 
to the gluon case, (\ref{eq:11}) and (\ref{eq:12}) do not contain
any large logarithmic factor $\log(\mu_{D}\rho_{eff})$
(here $\rho_{eff}$ is the typical transverse scale) 
which one could expect there if one would neglect the $\rho$-dependence
of $C(\rho)$.
 
The analytical formula (\ref{eq:12}) for the $N\!=\!1$ spectrum 
demonstrates that due to the finite-size effects at large energies, when 
$L_{f}^{\gamma}\gsim L$, the induced spectrum is strongly peaked 
at $x\approx 1$. As will be seen below
the $N\!=\!1$ dominates for RHIC and LHC conditions, and the spectrum with the 
$N\ge 2$ scatterings included remains sharply peaked at $x$ close to unity.
Since the $x$-integral on the right-hand side of (\ref{eq:1})
is also dominated by large $x$
one may expect that the induced photon emission to be an important
source of the direct photons.
It should be emphasized that our $1/(1-x)$ spectrum
does not mean enhancement of radiation at large $x$. It originates simply from
suppression of photon emission at small $x$ due to the finite-size effects.

Now let us discuss the influence of the induced gluon emission on
the photon bremsstrahlung ignored in the above derivation.
The gluon emission ,i.e., processes like 
$q\rightarrow \gamma g q$, and multi-gluon radiation, leads to quark 
energy loss,
and should reduce radiation of hard photons.
It is important that for soft gluons, dominating the quark energy loss, 
the gluon formation 
length, 
$L_{f}^{g}\sim \frac{2Ex(1-x)}{m_{q}^{2}x^{2}+m_{g}^{2}(1-x)}$, 
turns out to be of the same order as that for photons. 
For this reason, as was already noted, photon and 
gluon emission should be treated on an even footing.
An accurate analysis with gluon emission is a 
complicated task which is far beyond the scope of the present paper.
We account for influence of the induced gluon emission at qualitative level.
In line with the prescription suggested in \cite{BDMS_quenching}
for hadron spectra, we replace 
in (\ref{eq:1}) the cross section of quark production by an effective 
cross section
\beq
\frac{d\sigma_{q,eff}^{AA}(p_{T})}{dydp_{T}^{2}}=N_{bin}
\int \frac{dx}{(1-x)^{2}}\cdot \frac{dI(x,p_{T}/(1-x))}{dx}
\cdot\frac{d\sigma_{q}^{pp}(p_{T}/(1-x))}{dydp_{T}^{2}}\,,
\label{eq:13}
\eeq
where $dI(x,E)/dx$ is the probability distribution in 
the quark energy loss, and
$N_{bin}$ is the number of the binary nucleon-nucleon collisions.
In the $p_{T}$-region of interest the cross section of quark production
in $pp$ collisions can be parametrized as \cite{FMS}
\beq
\frac{d\sigma_{q}^{pp}(p_{T})}{dydp_{T}^{2}}\approx\frac{A}{(p_{T}+p_{0})^{n}}
\label{eq:14}
\eeq
with $p_{0}\approx 1.6$ GeV, $n\approx 8$ for RHIC, and
$p_{0}\approx 0.6$ GeV, $n\approx 5.3$ for LHC, 
the normalization constant $A$ will not
be important to us.
For such a $p_{T}$-dependence to good accuracy one can write the 
effective cross section of quark production
as
\beq
\frac{d\sigma(p_{T})_{q,eff}^{AA}}{dydp_{T}^{2}}\approx N_{bin}
\frac{d\sigma_{q}^{pp}(p_{T})}{dydp_{T}^{2}}\cdot
\int dx(1-x)^{n(p_{T})-2}\cdot \frac{dI(x,p_{T})}{dx}\,,
\label{eq:15}
\eeq
where
$$
n(p_{T})=-\frac{d}{d\ln p_{T}}\ln 
\frac{d\sigma_{q}^{pp}(p_{T})}{dydp_{T}^{2}}=\frac{n p_{T}}{(p_{T}+p_{0})}\,.
$$
Then the effective fragmentation function for photon radiation
accounting for gluon emission can be written in the form  
\beq
D_{q\rightarrow \gamma}^{eff}(x,E)=S_{g}(E)
\left[\frac{dP_{vac}(x,E)}{dx}+
\frac{dP_{ind}(x,E)}{dx}\right]\,,
\label{eq:16}
\eeq
where $S_{g}(E)$ is the gluon suppression factor given by
$$
S_{g}(E)\approx P_{0}(E)+
\int_{x_{min}}^{1} dx(1-x)^{n(E)-2} \frac{dI(x,E)}{dx}\,.
$$
Here we separated the probability of photon emission without gluons, 
$P_{0}$. In terms of the gluon spectrum, $dP_{g}/dx$, it reads 
$$P_{0}(E)=\exp\left(-\int_{x_{min}}^{1}dx 
\frac{dP_{g}(x,E)}{dx}\right)$$
with $x_{min}\sim m_{g}/E$. In numerical calculations we take 
$x_{min}=2m_{g}/E$.

We take the spectrum
in the radiated energy entering (\ref{eq:13}) in the form
\beq
\frac{dI(x,E)}{dx}=\exp\left(-\int_{x}^{1}dy 
\frac{dP_{g}(y,E)}{dy}\right)
\cdot\frac{dP_{g}(x,E)}{dx}\,.
\label{eq:17}
\eeq
The formula (\ref{eq:17}) is similar to the electron energy loss spectrum 
derived in \cite{Z_SLACSPS}. It works well for small energy loss 
$\Delta E\ll E$.
For RHIC and LHC the parametrization (\ref{eq:17}) 
reproduces the energy loss spectrum evaluated assuming independent gluon
radiation with accuracy $\sim 10-40$\%. It is enough to make qualitative
estimate of the effect of gluon emission on the photon spectrum.
An accurate calculation of the gluon suppression factor in the 
approximation of 
independent gluon emission  \cite{BDMS_quenching,GLV} does not make
sense because this
approximation itself does not have any serious theoretical justification.
It should be noted that the suppression of hadron spectra
evaluated using the energy loss spectrum (\ref{eq:17}) agrees well with that
observed at RHIC. Thus, our approximation in some sense is justified 
by the experimental data.

Neglecting the ISI effects the nuclear modification factor defined as
\beq
R_{AA}(p_{T})=\frac{1}{N_{bin}}\cdot
\frac{d\sigma_{\gamma}^{AA}(p_{T})/dydp_{T}^{2}}
{d\sigma_{\gamma}^{pp}(p_{T})/dydp_{T}^{2}}
\label{eq:18}
\eeq
can be approximately written as
\beq
R_{AA}(p_{T})\approx
S_{g}(p_{T})\cdot\frac{
\int_{0}^{1} dx x^{n(p_{T})-2}
\left[{dP_{vac}(x,p_{T})}/{dx}+
{dP_{ind}(x,p_{T})}/{dx}\right]
}
{\int_{0}^{1}  dx x^{n(p_{T})-2}
{dP_{vac}(x,p_{T})}/{dx}}\,.
\label{eq:19}
\eeq

We take the vacuum distribution in the form
\beq
\frac{dP_{vac}(x,E)}{dx}=
\frac{\alpha}{4\pi x}(4-4x+2x^{2})\cdot \int_{0}^{k_{max}^{2}}
dk^{2}\frac{k^{2}}{(k^{2}+\epsilon^{2})^{2}}\,,
\label{eq:20}
\eeq
where $\epsilon=m_{q}x$, and $k_{max}\approx E\max(x,(1-x))$.

To evaluate the gluon suppression factor $S_{g}$ we use the
formula for gluon spectrum similar to (\ref{eq:3}). However, we take
the thickness equals $L/2$. It seems to be a reasonable choice   
to account for the fact that for $L_{f}^{g}\sim L_{f}^{\gamma}$
about half of gluons are radiated after 
photon emission and cannot affect the photon bremsstrahlung.
Note that this reduces the suppression effect of gluon emission
as compared with the analyses \cite{DH,JOS2,JOS1,JJS}.

One remark is in order in connection with the above treatment of the gluon
effects. Our gluon suppression factor includes only the induced gluon
radiation. As far as the ordinary vacuum hard gluon radiation is concerned,
we assume that, due to small formation length, the corresponding 
suppression factors are approximately the same for $AA$- and $pp$-collisions.
For this reason one may ignore the vacuum gluon radiation in evaluating
the nuclear modification factor (if one uses the leading order vacuum spectrum 
(\ref{eq:20})).

\vspace{.2cm}
\noindent {\bf 3}. For numerical calculations we use the one-loop running 
coupling constant 
with $\Lambda_{QCD}=0.3$ GeV frozen at the value $\alpha_{s}=0.7$.
This parametrization is motivated by wanting to satisfy the relation
\beq
\int_{\mbox{0}}^{\mbox{\small 2 GeV}}\!dk\frac{\alpha_{s}(k)}{\pi}
\approx 0.36 \,\,
\mbox{GeV}\,
\label{eq:2}
\eeq  
obtained from the analysis of the heavy quark energy loss \cite{DKT}.  
To fix the quark and Debye screening mass we use the results of the
analysis within the quasiparticle model \cite{LH} of the lattice results.
For the relevant range of temperature of the plasma $T\sim (1-3)T_{c}$ 
($T_{c}\approx 170$ MeV is the
temperature of the deconfinement phase transition) the analysis 
\cite{LH} gives for the quark and gluon quasiparticle masses
$m_{q}\approx 0.3$ and $m_{g}\approx 0.4$ GeV.
With the above value of 
$m_{g}$ from the perturbative relation $\mu_{D}=\sqrt{2}m_{g}$ one
obtains $\mu_{D}\approx 0.57$ GeV.
We assume the Bjorken \cite{Bjorken} longitudinal expansion of the QGP with
$T\tau^{3}=T_{0}\tau^{3}_{0}$. For the initial conditions we use the
values suggested in \cite{FMS}
$T_{0}=446$ MeV and $\tau_{0}=0.147$ fm for RHIC, and
$T_{0}=897$ MeV and $\tau_{0}=0.073$ fm for LHC.
For RHIC the above condition were obtained from 
the charged particle pseudorapidity density 
$dN/dy\approx 1260$ measured by the PHOBOS experiment
\cite{PHOBOS}
assuming an isentropic expansion and rapid thermolization at 
$\tau_{0}\sim 1/3T_{0}$.
The LHC parameters correspond to $dN/dy\approx 5625$ estimated in 
\cite{KMS}. Note that, since the dominating $\rho$-scale in (\ref{eq:3})
$\propto \sqrt{z}$ for $z\ll L_{f}^{\gamma}$,
our results are not very sensitive to $\tau_{0}$.
For the upper limit of the $z$-integration in (\ref{eq:2}) we take 
$L=R_{A}\approx 6$ fm
\footnote{For our choice of the initial conditions the life-time of QGP is 
$\sim 3$ fm for RHIC. However,  
in the interval $\tau\sim 3-6$ fm the density of the
mixed phase is practically the same as that for the pure QGP phase.}. 
This seems to be a reasonable value for central heavy-ion collisions
since  
due to the transverse expansion the hot QCD matter
should cool quickly at $\tau\gsim R_{A}$ \cite{Bjorken}.

In Figs.~2 and 3 we show the $x$-dependence of the probability distribution
of the induced $q\rightarrow \gamma q$ transition (solid line) 
for several quark energies for RHIC and LHC.
For comparison the vacuum spectrum (dashed line) is also shown.
One sees that in the region of large $x$, which dominates 
the $x$-integrals on the right-hand side of Eq.~(\ref{eq:19}), the 
induced radiation exceeds the vacuum one (especially for the LHC case).
To illustrate the influence of the LPM effect on the induced photon
emission we also show  
the contribution of the $N\!=\!1$ scattering (long-dashed line). It is seen 
that for $x\sim 0.6-0.8$
the LPM effect reduces the induced bremsstrahlung by a factor of 
$\sim 0.8$ for RHIC and $\sim 0.5$ for LHC. 
Note that the LPM effect and finite-size effects  diminish in strength 
as $x\rightarrow 1$ since $L_{f}^{\gamma}$ becomes small and the spectrum
should be close the Bethe-Heitler one in this limit.
To demonstrate the role of the finite kinematic limits (neglected in 
(\ref{eq:3})) in Figs.~2,~3 we also show the $N\!=\!1$ scattering contribution 
evaluated using the set of diagrams shown in Fig.~1 
with finite kinematic limits (dotted line). 
One can see that the kinematic effects 
become important at $x$ close to unity. Numerical calculations show that
they reduce the integral over $x$ in the numerator
in (\ref{eq:19}) by $\sim 20-30$\%. 
However, this suppression to good accuracy is compensated by 
the increase in the gluon suppression factor
due to similar kinematic effects for gluon emission. For this reason
we neglect the kinematic effects in calculation 
of the nuclear modification factor (\ref{eq:19}).

In Fig.~4 we plot results for the $p_{T}$-dependence 
of the nuclear modification factor (\ref{eq:19}). We also show the results
without gluon suppression factor (dashed line), i.e., for $S_{g}=1$.  
One sees that despite strong suppression 
due to gluon emission we obtain $R_{AA}\gsim 1$ at $p_{T}\sim 5-15$ GeV. 
This means that the FSI effects can 
enhance the direct photon production (especially for LHC energies, where  
bremsstrahlung is the dominating partonic mechanism). Of course, one should 
bear in mind that uncertainties in our theoretical predictions can 
be about $30\div 50$\%. Nevertheless, one can say that 
it is practically excluded that in central $AA$-collisions the direct photon 
production is strongly suppressed due to quark energy loss, as was 
predicted in \cite{DH,JOS2,JOS1,JJS}.

\bigskip
\noindent {\large \bf Acknowledgements.} 
I am grateful to S.~Peigne for discussions. 
I thank especially P.~Aurenche for discussions and useful comments
on the paper. I am also grateful to the LAPTH and the High Energy Group of 
the ICTP for the kind hospitality during my
visits to Annecy and Trieste where a part of this work was done.

\newpage

%------------------------------------------------------------------
\begin{center}
{\Large \bf Figures}
\end{center}
%------------------------------------------------------------------

\begin{figure}[h]
\begin{center}
\epsfig{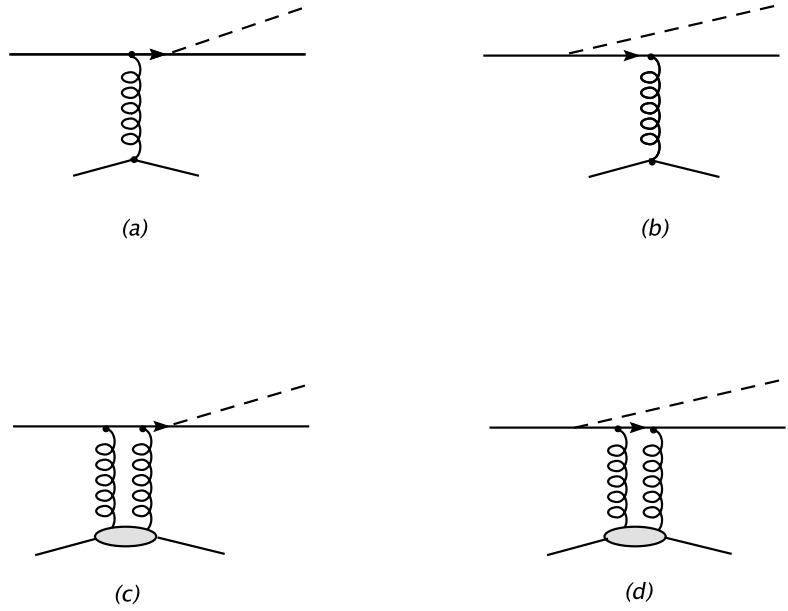}
\end{center}
\caption[.]{
The set of Feynman diagrams corresponding to the $N\!=\!1$ scattering
induced spectrum.
}
\end{figure}

\begin{figure}[h]
\begin{center}
\epsfig{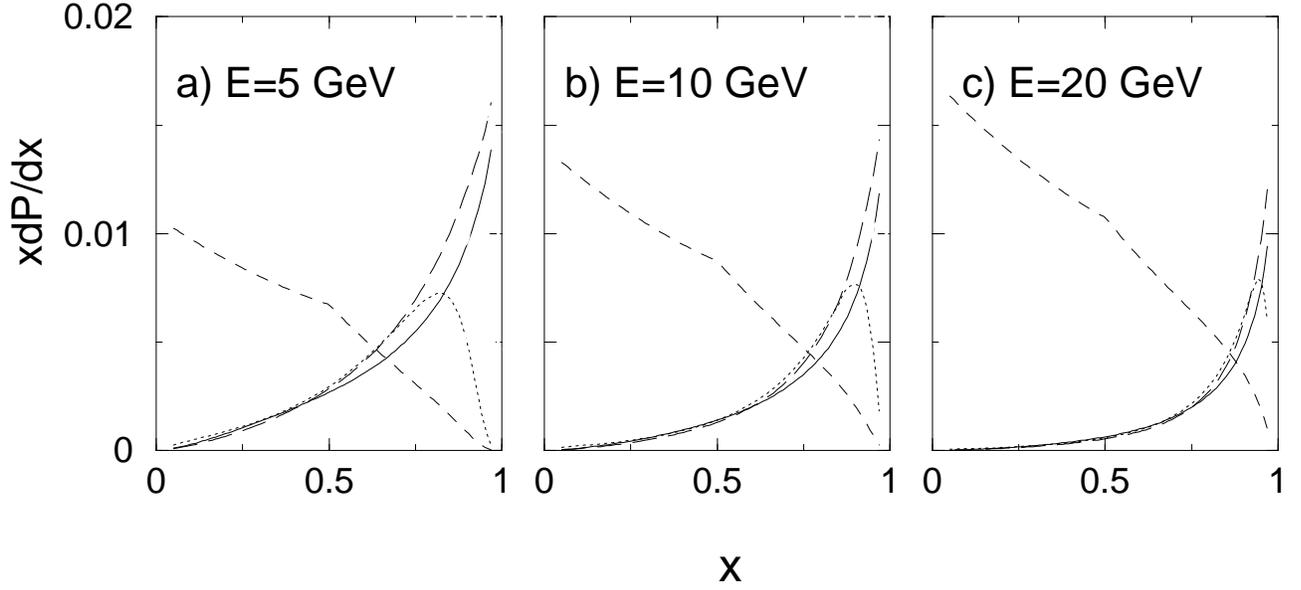}
\end{center}
\caption[.]{
The photon spectrum of the $q\rightarrow \gamma q$ transition
for RHIC conditions. 
The solid line shows contribution from the induced photon emission
calculated using the formula (\ref{eq:3}).
The dashed line shows the vacuum spectrum (\ref{eq:20}).
The $N\!=\!1$ scattering contribution
to the induced spectrum is shown by 
the long-dashed curves for infinite kinematic boundaries, and
by the dotted curves for finite kinematic boundaries.
}
\end{figure}

\begin{figure}[h]
\begin{center}
\epsfig{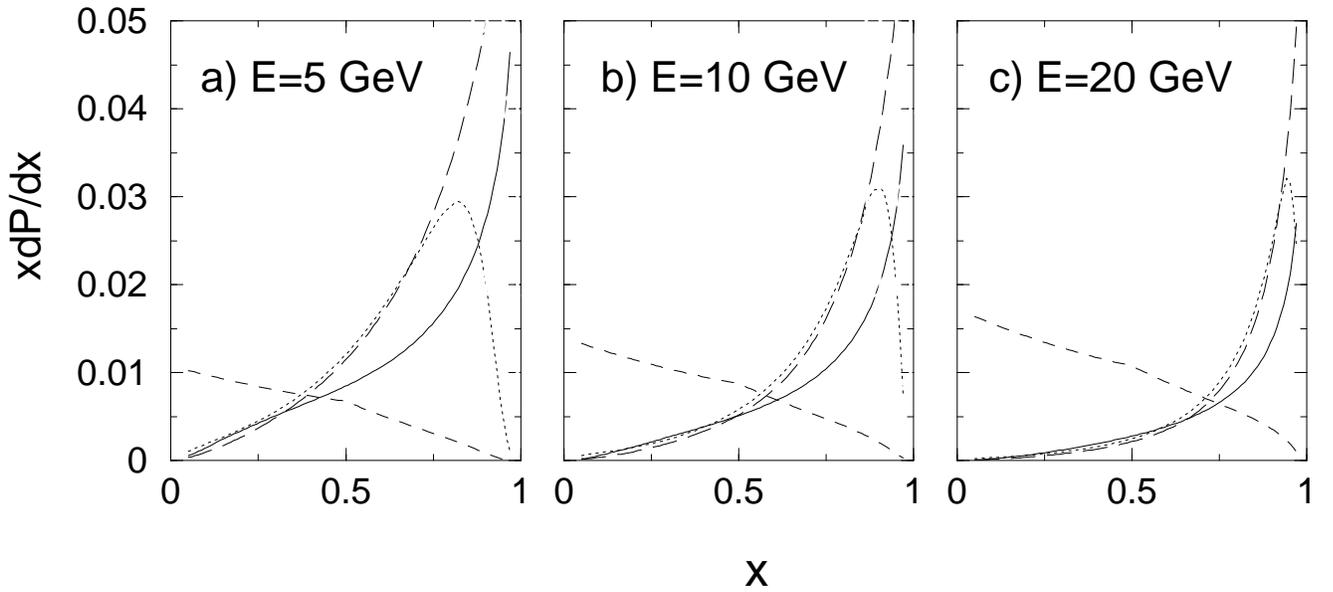}
\end{center}
\caption[.]{
The same as in Fig.~2 but
for LHC.
}
\end{figure}

\begin{figure}[t]
%\begin{center}
\epsfig{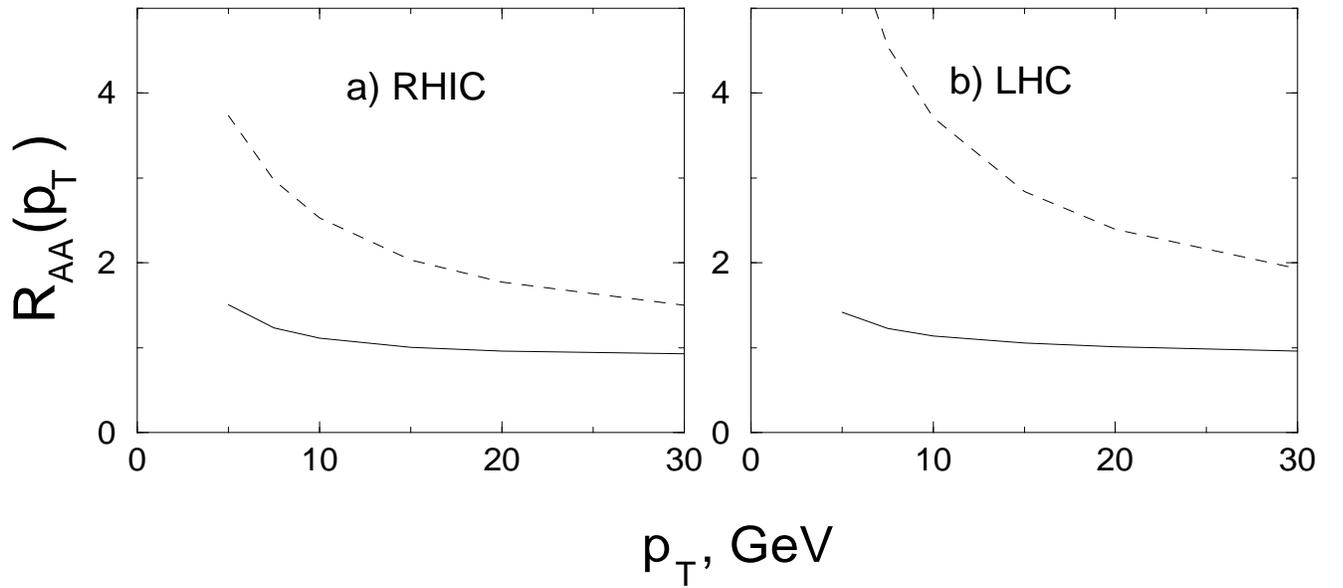}
%\end{center}
\caption[.]{
The $p_{T}$-dependence of the nuclear modification factor (\ref{eq:19})
(solid line) for RHIC (a) and LHC (b) conditions. The dashed line
shows the results without gluon suppression factor $S_{g}$ on the right-hand
side of (\ref{eq:19}).
}
\end{figure}

\end{document}